\documentclass[prd,groupedaddress, showpacs, twocolumn, amsmath,amssymb,nofootinbib]{revtex4-1}

\usepackage{graphicx}
\usepackage{dcolumn}
\usepackage{bm}
\usepackage{amsmath,graphics,color}
\usepackage{color,psfrag}
\usepackage{mathrsfs}

\newcommand{\beq}{\begin{equation}}
\newcommand{\eeq}{\end{equation}}
\newcommand{\bea}{\begin{eqnarray}}
\newcommand{\eea}{\end{eqnarray}}

\def\d{\partial}

\def\d{\delta}

\def\phi{\varphi}

\def\l{\lambda}
\def\m{\mu}
\def\n{\nu}

\def\s{\sigma}

\def\th{\theta}

\begin{document}

\title{Reversible and Irreversible Spacetime Thermodynamics\\
for General Brans-Dicke Theories }

\author{Goffredo Chirco}
\author{Christopher Eling}
\author{Stefano Liberati}
\affiliation{SISSA, Via Bonomea 265, 34136 Trieste, Italy and INFN Sezione di Trieste, Via Valerio 2, 34127 Trieste, Italy}

\date{\today}

\begin{abstract}
We derive the equations of motion for Palatini $F(\mathcal{R})$ gravity by applying an entropy balance law $TdS=\d Q+\d N$ to the local Rindler wedge that can be constructed at each point of spacetime. Unlike previous results for metric $F(R)$, there is no bulk viscosity term in the irreversible flux $\d N$. Both theories are equivalent to particular cases of Brans-Dicke scalar-tensor gravity. We show that the thermodynamical approach can be used \emph{ab initio} also for this class of gravitational theories and it is able to provide both the metric and scalar equations of motion. In this case, the presence of an additional scalar degree of freedom and the requirement for it to be dynamical naturally imply a separate contribution from the scalar field to the heat flux $\delta Q$. Therefore, the gravitational flux previously associated to a bulk viscosity term in metric $F(R)$ turns out to be actually part of the reversible thermodynamics. Hence we conjecture that only the shear viscosity associated with Hartle-Hawking dissipation should be associated with irreversible thermodynamics.\\

\end{abstract}
\maketitle

\section{Introduction}

In the past decade, there has been significant interest in the idea of gravity as the thermodynamics of a quantum theory associated to some underlying microscopic structure of spacetime \cite{Jacobson:1995ab,Padmanabhan:2009vy,Eling:2006aw,Eling:2008af,Chirco:2009dc,others}. The first striking hint in this direction was given by Jacobson \cite{Jacobson:1995ab}, who derived the Einstein equation as an equilibrium equation of state, starting from the thermodynamical properties of the vacuum. The key ingredients of the derivation were the possibility to characterize the Minkowski vacuum perceived by an uniformly accelerated observer as a canonical thermal state (the Unruh effect), together with the idea that the local acceleration horizon, associated with the former observer, could be analogous to a tiny piece of a black hole event horizon, thereby allowing for an area-entropy proportionality assumption.

In this context, the coupling between geometry and matter was then provided by demanding a local equilibrium Clausius relation, $TdS= \d Q$, as the main tool to relate the variation of entropy of the vacuum fields $dS$ (horizon area deformation) to the perturbative effects of a local flux of matter energy $\d Q$ through the acceleration horizon.

More recently, in the attempt to understand whether and how the thermodynamical derivation of the Einstein equation could be generalized to allow for the higher curvature terms expected by the effective field theory, it was realized that the thermodynamical approach can be naturally generalized from an equilibrium to a non-equilibrium thermodynamical setting \cite{Eling:2006aw,Eling:2008af}.
Indeed, it was pointed out in \cite{Eling:2006aw} that in a natural extension of the framework to $F(R)$ gravity theory, the horizon entropy proportionality to a function of the Ricci scalar necessarily led to a break down of the thermodynamical equilibrium. In order to recover the $F(R)$ gravity field equation from the thermodynamical prescription, it was necessary to generalize the local equilibrium condition to a more general entropy balance law. The new local equilibrium condition, $T dS=\d Q+\d N$ (the generalized Clausius relation), contains extra irreversible entropy production terms due to the source contributions for the horizon area/entropy evolution which are quadratic in expansion and shear of the null geodesic bundle comprising the horizon.

Already interpreted in \cite{Eling:2006aw} as dissipative effects, these terms were linked to purely gravitational/internal degrees of freedom of the theory \cite{Chirco:2009dc}. In particular, the entropy production terms associated with the quadratic shear were recognized to be the equivalent, for the local horizon setting, of the tidal heating term associated to dissipation of the black hole horizon's perturbations via gravitational fluxes. Similarly, the entropy production terms associated with the quadratic expansion seemed to be an extra purely scalar dissipative contribution.

However, the scalar dissipative contribution seems to be formally local, as a derivative of a scalar field at a point. Being local, this term would be frame independent, thereby it would exist for any observer (accelerated or inertial) in the local patch of spacetime and would always end up describing the dynamics of the global spacetime. Therefore, an interpretation of this term as a dissipative contribution would imply waves of the scalar field would be dissipative in the spacetime. This is inconsistent with the fact that classical gravitational theories are time reversal invariant.

In order to address this problem, we start in this work by extending the thermodynamical approach, previously applied to metric $F(R)$ gravity, and work in the Palatini formalism, where the connection is a priori an independent variable from the metric. In this theory, the connection is not a propagating degree of freedom and can be eliminated by an algebraic relation. As a consequence, unlike the metric $F(R)$ case, we show that no extra entropy production terms is required in order to get the equations of motion from the generalized local equilibrium condition. This is in agreement with what one would expect from the reasoning proposed in \cite{Chirco:2009dc} regarding the interpretation of these terms as linked to the dynamical gravitational degrees of freedom of the theory.

However, both metric and Palatini $F(R)$ gravity are equivalent, at the classical level, to Brans-Dicke theories with particular values of the Dicke constant. With this motivation, we then consider an entropy density that is a scalar spacetime function, promoting the inverse of the Newton constant to be an independent scalar field. In this way, we can show that the thermodynamic derivation in this case can capture both the field equations of the metric and the scalar field for these theories.

In this scalar-tensor framework, the extra entropy contribution associated with the quadratic expansion in the metric $F(R)$ appears as a separate contribution from the scalar field to the reversible heat flux $\d Q$. Therefore, the gravitational flux previously associated to a bulk viscosity term in metric $F(R)$ is understood as a component of the reversible sector in the thermodynamical formalism. In this picture, we are able to generalize the thermodynamical approach to general Brans-Dicke scalar-tensor theories of gravity.

We start in section \ref{mfr} with a review of the spacetime thermodynamics approach developed for Einstein gravity in and its extension to metric $F(R)$ theory. In section \ref{pal}, we reconsider the thermodynamical formalism for Palatini gravity. We show how the equilibrium condition, which was just an auxiliary condition in the metric formalism, now implies the field equation associated with the variation of the Lagrangian with respect to the independent connection. Then, we derive the equation of motion from the local equilibrium assumption, where no additional bulk viscosity term is needed in the analysis. In section \ref{str}, the interpretation of the bulk viscosity term is reconsidered through the equivalence of metric and Palatini versions of $F(R)$ to particular cases of scalar tensor theories. This provides the background to generalize the argument to a general Brans-Dicke theory in section \ref{GeneralST}. We conclude with a summary and discussion of future work.

\section{Thermodynamics of spacetime and metric $F(R)$ gravity} \label{mfr}

Here we will first review the thermodynamics of spacetime approach originally developed for Einstein gravity \cite{Jacobson:1995ab} and its extension to metric $F(R)$ gravity \cite{Eling:2006aw} (also see \cite{Chirco:2009dc}) as this will be the basis for our results in the following sections. The foundation of the approach is in the thermodynamical properties of the flat Rindler spacetime, which are independent of gravity.  In the Rindler coordinates $Y^\mu = (\tau, \xi, x,y)$, a flat manifold has the line element
\beq ds^2 = g_{\mu \nu} dY^\mu dY^\nu = \kappa^2 \xi^2 d\tau^2- d\xi^2 - dx^2- dy^2. \label{Rindlermetric} \eeq
Here $\kappa$ is an arbitrary constant with dimensions $[L]^{-1}$ (we work in units where $\hbar=c=k_B=1$) associated with the normalization of the timelike Killing vector $\partial_\tau$.  These coordinates only cover a ``wedge" subregion of the spacetime, in terms of global Minkowski coordinates the range where $z > |t|$. The timelike Killing flow $\chi^\mu = (\partial_{\tau})^\mu$ is equivalent to a continuous boost in the $z$ direction. The respective boost time parameter  $\tau$ is proportional to the proper time along the worldlines of a uniformly accelerated observer, defined by the $\xi=const$ hyperbolas. The surface $\xi=0$ is the null surface $z=t$, which acts similar to future event horizon of a black hole since the points ``inside" are causally disconnected from the accelerated observers.

When restricted to the Rindler wedge, the usual global Minkowski vacuum state $|0  \rangle $ in quantum field theory turns out to be equivalent to a Gibbs thermal state with an Unruh-Tolman temperature \cite{Unruh:1976db, Sewell}
\beq T = \frac{\kappa}{2 \pi}, \label{temp} \eeq
and a thermal entanglement entropy \cite{entanglement}
\beq S_{ent} = - Tr \rho \ln \rho. \eeq
As said, the factor $\kappa$ is an arbitrary rescaling factor for the proper time $\tau$, and as such it can always be set to one. Hence from this point forward we will usually work with the dimensionless ``temperature" $1/2\pi$. Remarkably, the entanglement entropy associated with the fields in the Rindler wedge scales not like the volume of the wedge, but instead with the area of the Rindler horizon boundary, just like the Bekenstein-Hawking entropy of a black hole.  On the other hand, in quantum field theory it is also quadratically ultraviolet (UV) divergent and we will assume a regulator has been introduced in order to render it finite.

Now a general spacetime is a curved manifold, but the equivalence principle implies that the neighborhood of any point $p$ is approximately flat. In this flat patch of spacetime, one can always construct a local Rindler wedge system. Let us be more precise about this construction. The key quantity we need is a localized version of the boost Killing vector $\chi^\mu$, vanishing at $p$. This object will allow us to define a local temperature and eventually a notion of heat flow. There are no Killing vectors for an arbitrary spacetime, but Killing's equation $2 \nabla_{(\mu} \chi_{\nu)} = 0$ can be solved locally around $p$ up to $O(x^3)$ using Riemann normal coordinates ${x^\mu}$, where
\beq g_{\mu \nu} = \eta_{\mu \nu} + O(x^2). \eeq

\begin{figure}[h!]
\includegraphics[width=\columnwidth]{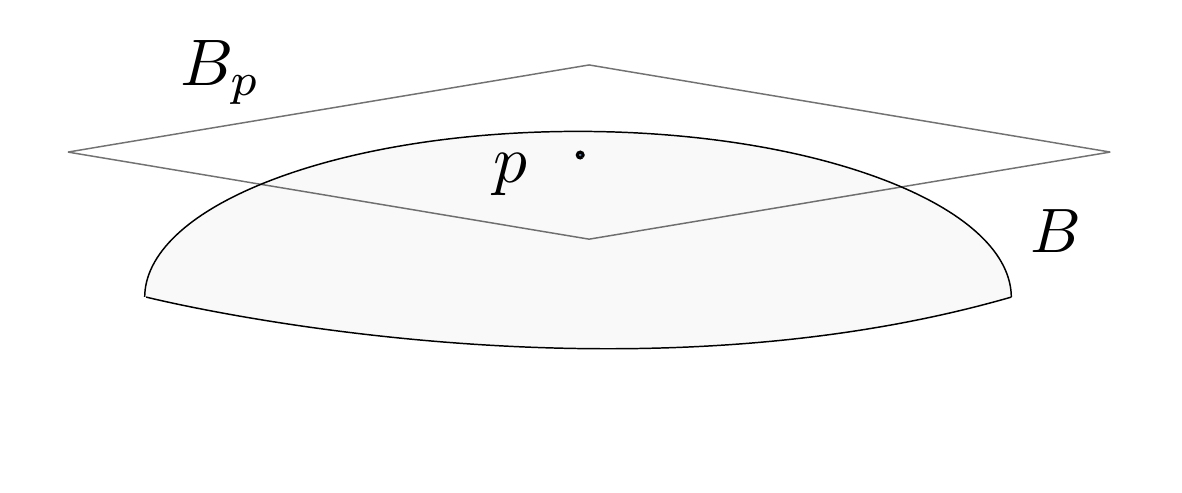}
\caption{The 2-surface $B$ agrees at $p$ with the tangent plane $B_p$  preserved by the flow of the local Killing vector. }\label{plane}
\end{figure}

In the neighborhood of each point in spacetime we think of the thermal system as the local Minkowski vacuum $|0  \rangle $ restricted to the local Rindler wedge. Note that the full Minkowski patch, where there is no local horizon and no coarse graining, has zero temperature and entropy and is not the thermal system in question. To probe the behavior of this system out of equilibrium one constructs a local causal horizon at $p$ as follows. Choose a spacelike 2-surface patch $B$ including $p$, and choose one side of the boundary of the causal past of $B$. Near $p$,  this boundary is a congruence of null geodesics orthogonal to $B$. These comprise the horizon.  At $p$, $B$ agrees with the tangent plane $B_p$ preserved by the flow of the local Killing vector (Fig. \ref{plane}). \\

We arrange $\chi^\mu$ so that it is future pointing on the causal horizon. Therefore on the horizon generator through $p$ we have (up to the ambiguities at $O(x^3)$) $\chi^\mu = -\lambda k^\mu$, where $k^\mu$ is the affinely parameterized null tangent vector $k^\mu$. This affine parameter obeys the formula
\beq \lambda = -e^{-v}, \label{affineKilling} \eeq
where $v$ is the ``Killing time" parameter defined via $\chi^\mu \nabla_\mu v = 1$ characterizing the time flow of the local wedge system.

The choice of how $B$ is warped in the spacetime determines the horizon expansion and shear, and therefore the deviation of the system from equilibrium. Using (\ref{affineKilling}) one can show the relationship between the expansion and shear of the local horizon defined in terms of $v$ and corresponding quantities in $\lambda$ is
\beq \hat{\theta}=- \lambda\, \theta ~~ \hat{\sigma}=- \lambda\, \sigma. \eeq
The Killing quantities automatically vanish because $p$ is a fixed point of the local boost flow, consistent with our notion of local thermodynamic equilibrium there. Whether the affine quantities are zero or not in turn determines the rate at which this equilibrium is approached. For example, when $\sigma_{|p} = 0$, $\hat{\sigma} \sim e^{-2v}$, while for $\sigma_{|p} \neq 0$, $\hat{\sigma} \sim e^{-v}$.

The basic idea is to impose a general entropy balance law on this local system.  When the thermal density matrix $\rho$ associated with the quantum fields in the wedge is perturbed, the change in the entanglement entropy should be related to the change in mean energy via the entropy balance law
\beq dS = \d Q/T+\delta N/T. \label{entropybalance}\eeq
We assume the change in the mean energy of the system is due to a flux into the unobservable region of spacetime, which is perfectly thermalized by the horizon system and therefore can be thought of as heat. It is assumed $\d Q$ is the flow of boost matter energy current  $T^{M}_{\mu \nu} \chi^\nu$ across the horizon,  in terms of the affine $k^\mu$ this is
\beq \frac{\d Q}{T} = 2\pi \int T^{M}{}_{\m \n} k^\m
k^\n (-\l) d\l \sqrt{h} d^2 x, \label{dQ/T} \eeq
%
where $\sqrt{h}$ is the cross-sectional area element on the causal horizon.

The additional $\delta N$ is an irreversible internal entropy production term, or uncompensated heat. This term arises in a slower approach to equilibrium.  Using linear constitutive relations between fluxes of momentum in a fluid and the thermodynamic ``forces" given by gradients of a fluid velocity, it was argued in \cite{Eling:2006aw} that the entropy production can be expressed in terms of the squared shear $\hat{\sigma}_{\mu \nu}$ and expansion $\hat{\theta}$ of the flow
\beq \frac{\delta N}{T} = \frac{2 \eta}{T} \hat{\sigma}_{\mu \nu} \hat{\sigma}^{\mu \nu} + \frac{\xi_B}{T} \hat{\theta}^2, \label{eq:unheat}\eeq
where $\eta$ and $\xi_B$ are shear and bulk viscosities respectively.

The entropy balance law (\ref{entropybalance}) we postulated requires a corresponding change in the entanglement entropy associated with the local Rindler wedge.  In \cite{Jacobson:1995ab} a constant, universal UV cut-off $\alpha$ with units of $[L]^{-2}$ was postulated to make the entropy density $s$ finite and the total entropy just proportional to the horizon cross-sectional area.  Imposing the entropy balance law ultimately yielded two results, which we will briefly summarize. The Einstein equations (with an undetermined cosmological constant and the Newton constant $G_N = (4 \alpha)^{-1}$) are associated with reversible changes in the global spacetime. This is consistent with the time reflection symmetry of Einstein's theory. In the irreversible sector, associated with the horizon, the shear viscosity was fixed to have the universal ratio $\eta/s = 1/4\pi$ \cite{Eling:2006aw, Eling:2008af}. The shear viscosity term coincides with the Hartle-Hawking tidal heating term and therefore appears to be associated with the purely tensorial gravitational degrees of freedom in the theory  \cite{Chirco:2009dc}.

However, one can imagine that the entanglement entropy density has a more complicated structure in general. For example, the only known theory of gravity consistent with the {\it strong} equivalence principle is Einstein gravity. The strong equivalence principle implies that gravity is purely geometrical. Physics (with gravity included) is the same in any locally flat region of spacetime, which means the UV cutoff is constant, $G_N$ is universal, and there are no extra gravitational fields.  However, in more general theories of gravity, the strong equivalence principle is not satisfied, suggesting one should consider an entanglement with a spacetime dependent cutoff and as a result, more general entropy functionals.

In \cite{Eling:2006aw} it was assumed the entropy density is a general function of the Ricci scalar $f(R)$, so that the total entropy is
\beq S = \alpha \int \sqrt{h} f(R) d^2 x, \eeq
In this case we will review the details of how the entropy balance law is imposed. The variation of this entropy in terms of the affine parameter $\lambda$ has the form
\beq \delta S = \alpha \int \sqrt{h} \left(f \theta + \frac{d f}{d\lambda}\right) d\lambda d^2 x. \label{metricdS} \eeq
To compare with the heat flux in (\ref{dQ/T}) we perform a series expansion in $\lambda$ around the point $p$,
\bea
\delta S &=& \alpha \int \sqrt{h} \left[\left(\theta f +  \frac{d f}{d\lambda}\right)\right.+ \nonumber \\
&+&\lambda \left.\left(\theta \frac{d f}{d\lambda} + f \frac{d \theta}{d\lambda}+\frac{d^2 f}{d\lambda^2} + f \theta^2 + \theta \frac{d f}{d\lambda}\right)\right]_p. \label{metricdS2}
\eea
The first three terms in the $O(\lambda)$ piece come from the derivative of the integrand in (\ref{metricdS}), while the last two terms are associated with the derivative of the transverse volume element $\sqrt{h}$.

We now match this expansion with the heat flux and irreversible heat at lowest orders in $\lambda$ as required by the entropy balance law. Because we are dealing with quantities evaluated at $p$, equating the integrals (\ref{metricdS2}) and (\ref{dQ/T}) is equivalent to equating the integrands. Since there is no zeroth order part of these fluxes (they must vanish at the equilibrium point $p$, see for example (\ref{dQ/T})), we have the equilibrium condition that
\beq \left(\theta f +  \frac{d f}{d\lambda}\right)_{p} = 0. \label{zerothorder}\eeq
One can always construct a 2-surface $B$ at $p$ to satisfy this condition. Now consider the $O(\lambda)$ terms of (\ref{entropybalance}). First, we replace $d\theta/d\lambda$ with the right hand side of the Raychaudhuri equation
\beq \frac{d\th}{d\l} = -\frac{1}{2}\th^2-\s_{\m \n} \s^{\m \n}-R_{\m \n}k^\m k^\n, \label{Ray}\eeq
Comparing the integrands, we find at $p$
\bea \alpha(-f R_{\m \n} k^\m k^\n + k^\m k^\n \nabla_\m \nabla_\n f - \frac{3}{2} f \theta^2- \phi \sigma_{\m \n} \sigma^{\m \n})   \nonumber \\
=
2\pi (-T^{M}{}_{\m \n} k^\m k^\n - 2\eta \s_{\m \n} \s^{\m \n} - \xi_B \theta^2). \label{matching} \eea
Note that in writing this equation we used $d/d\lambda = k^\mu \nabla_\mu$ and the affine geodesic equation $k^\mu \nabla_\mu k^\nu = 0$.
Using (\ref{zerothorder}), the expansion can be expressed as a kinetic term for $f$,
\beq \frac{3}{2} f \theta^2 = \frac{3}{2f} k^\m k^\n \nabla_\m f \nabla_\n f. \label{kinetic} \eeq

If we consider this kinetic term to be a part of the reversible sector and demand the equation hold at any point $p$ and for all null vectors $k^\mu$,  we get
\beq  f R_{\m \n} -  \nabla_\m \nabla_\n f + \frac{3}{2f} \nabla_\m f \nabla_\n f + \Phi g_{\mu \nu} = \left(\frac{2\pi}{\alpha}\right) T^{M}{}_{\m \n}, \label{testreversible}\eeq
where $\Phi$ is an arbitrary function. Local energy-momentum conservation requires $\nabla^\m T^{M}{}_{\m \n} = 0$ and therefore gives us a condition to solve for $\Phi$. However, as was shown in \cite{Eling:2006aw}, with the additional kinetic term (\ref{kinetic}) present, there is a contradiction and one cannot solve for $\Phi$. In this case, conservation of energy-momentum seems to require this term to contribute to the irreversible entropy production. Put differently, one must have the identifications
\bea
\eta &=& \frac{\alpha f}{4\pi} \\
\xi_B &=& \frac{3 \alpha f}{4\pi}.
\eea
to cancel out the kinetic and shear squared terms. Note that since $s= \alpha f$, the ratio $\eta/s = 1/4\pi$ still. The purely reversible changes are now associated with the equation
\beq  f R_{\m \n}  - \nabla_\m \nabla_\n f + \Phi g_{\mu \nu} = \left(\frac{2\pi}{\alpha}\right) T^{M}{}_{\m \n}, \label{metricrev} \eeq
Imposing local energy conservation and then using the contracted Bianchi identity and the commutator of covariant derivatives, one finds that $\Phi = \Box f - \frac{1}{2} F$, where $F(R)$ is the Lagrangian of the theory and defined here by the equation $f=dF/dR$.  As a result (\ref{metricrev}) now coincides with the equation of motion derived from the action
\beq I_{met}= \frac{\alpha}{4\pi} \int \sqrt{-g} d^4 x (F(R) + L_{matt}). \label{metricaction} \eeq

\section{Palatini $F(\mathcal{R})$ gravity} \label{pal}

In this section we will address whether the formalism of Palatini gravity is compatible with the thermodynamic approach. In the Palatini formalism, one treats the connection as an independent variable a priori. The Riemann tensor $\mathcal{R}_{\mu \nu \rho \lambda}$, constructed out of this connection, is now  therefore also independent of the metric.

Our starting point will be an entropy density which is an arbitrary function of the independent Ricci scalar
\beq S = \alpha \int \sqrt{h} f(\mathcal{R}) d^2 x. \eeq
If the thermodynamic approach works in this case, we expect now {\it two} equations, which are the equations of motion following from the variation of the Lagrangian
\beq I_P = \frac{\alpha}{4\pi}  \int \sqrt{-g} (F(\mathcal{R}) + L_{matt}(g_{\mu \nu}, \psi)) \label{Palatiniaction} \eeq
with respect to $g^{\mu \nu}$ and $\Gamma^{\lambda}{}_{\mu \nu}$. Note that in the Palatini formalism the matter part of the action is assumed not to depend on the independent connection.

Palatini $F(\mathcal{R})$ gravity theory has been discussed extensively over the past decade as an alternative theory of gravity \cite{Sotiriou:2008rp}. Here we pause briefly to review the properties of this theory. Defining $f = dF/d\mathcal{R}$ as before, the equations of motion are
\beq
f(\mathcal{R}) \mathcal{R}_{\mu \nu} - \frac{1}{2} F(\mathcal{R}) g_{\mu \nu}=\left(\frac{2\pi}{\alpha}\right) T_{\mu \nu} \label{Palatinimetric}
\eeq
\beq
 \bar{\nabla}_{\sigma}(\sqrt{-g} f(\mathcal{R}) g^{\sigma(\mu} {\delta^{\nu)}}_{\lambda} - \bar{\nabla}_{\lambda}(\sqrt{-g} f(\mathcal{R}) g^{\mu \nu})=0 , \label{Palatiniconnection}
\eeq
where $\bar{\nabla}$ represents the covariant derivative defined with respect to the independent connection. The connection equation is equivalent to the more compact condition
\beq \bar{\nabla}_{\lambda}(\sqrt{-g} f(\mathcal{R}) g^{\mu \nu}) = 0. \label{metriccomp} \eeq
Note that when $f$ is equal to a constant this equation reduces to the usual metric compatibility condition for $g^{\mu \nu}$. Therefore we see the textbook equivalence of the Palatini and metric formalisms of GR. In general however, (\ref{metriccomp}) implies the conformally related metric
\beq \bar{g}_{\mu \nu}= f(\mathcal{R}) g_{\mu \nu} \eeq
is compatible with the connection.  Imposing this condition, one can relate the Ricci tensor and scalar constructed from $\bar{g}_{\mu \nu}$ to the metric quantities
\bea
\mathcal{R}_{\mu \nu} &=& R_{\mu \nu} + \frac{3}{2} \frac{1}{f^2} (\nabla_\mu f')(\nabla_\nu f)  \nonumber \\ &&
- \frac{1}{f}(\nabla_\mu \nabla_\nu- \frac{1}{2} g_{\mu \nu} \Box ) f \label{RicciTconform}\\
\mathcal{R} &=& R + \frac{3}{2} \frac{1}{f^2} (\nabla_\mu f)(\nabla^\mu f)+ \frac{3}{f} \Box f. \label{RicciSconform}\eea
These equations can be substituted into (\ref{Palatinimetric}) to yield \cite{Sotiriou:2008rp}
\bea
f G_{\mu \nu} &=& \left(\frac{2\pi}{\alpha} \right) T_{\mu \nu} - \frac{1}{2} g_{\mu \nu} \left(f \mathcal{R}-F(\mathcal{R}) \right)  \nonumber\\ &&
+ \frac{1}{f}(\nabla_\mu \nabla_\nu - g_{\mu \nu} \Box) f - \frac{3}{2} \frac{1}{f^2} [(\nabla_\nu f)(\nabla_\nu f) \nonumber \\ &&
 - \frac{1}{2} g_{\mu \nu} (\nabla f)^2]. \label{reducedeom}
\eea
Solving the trace of (\ref{Palatinimetric}) for $\mathcal{R}$ in terms of $T$,  one can completely eliminate the connection as an independent variable and reduce the system to one equation of motion that looks like GR with a modified source. Following the reasoning of \cite{Chirco:2009dc} one then should expect that no bulk viscosity term appears in this case, as no-extra dynamical gravitational degree of freedom with respect to the metric appears in Palatini $F(\mathcal{R})$ gravity.

\subsection{Thermodynamic formalism}

With the connection now an independent variable and the metric no longer a priori compatible with it, we want to consider the effect (if any) on the basics of the thermodynamics of spacetime formalism introduced in Section II. This is worth doing not only for checking the validity of the above discussed expectation (no-bulk viscosity associated to $F(\mathcal{R})$) but also as a first step towards the generalization of the spacetime thermodynamics formalism to the broader class of metric-affine theories of gravity.

In the neighborhood of each point, spacetime is still locally flat and we still can construct the boost Killing vector $\chi^\mu$ since the Killing equation $\mathcal{L}_\chi g_{\mu \nu} = 0$ does not depend on the connection. On the local horizon $\chi^\mu$ is still a null generator. However,  in the presence of an independent connection, there is a priori an ambiguity in whether $\chi^\mu$ is a geodesic with respect to the independent connection or the metric one. We have either
\beq \chi^\nu \bar{\nabla}_\nu \chi^\mu = \bar{\kappa} \chi^\mu \label{indgeodesic} \eeq
or
\beq \chi^\nu \nabla_\nu \chi^\mu = \kappa \chi^\mu, \label{metricgeodesic} \eeq
and the corresponding choices for the affine parametrization
\bea \chi^\mu &=& -\bar{\kappa} \bar{\lambda} \ell^\mu \\
\chi^\mu &=& -\kappa \lambda k^\mu. \eea
Each vector above is affinely parameterized with respect to either the independent or the Levi-Civita connection: $\ell^\nu \bar{\nabla}_\nu \ell^\mu = 0$ or $k^\nu \nabla_\nu k^\mu = 0$. Therefore,  in the entropy change $\delta S$, one must consider changes with respect to either affine parameter. A priori the different Clausius relations could yield two different equations of motion. In the next two subsections we will consider variations with respect to $\bar{\lambda}$ and $\lambda$ in turn.

\subsubsection{Variation using independent connection}
\label{Generalvariation}

First we consider the heat flux. We express the Killing field in terms of the affine $\ell^\mu$ using the formula $\chi^\mu = -\bar{\lambda} \ell^\mu$, where we scale the $\bar{\kappa} = 1$ as usual.\footnote{It is worth noticing here that there is an ambiguity about which $\kappa$ (barred or unbarred) would actually appear in the Tolman-Unruh temperature $T$, see Eqn. (\ref{temp}). If the unbarred $\kappa$ is chosen, due to the coupling of matter fields only to metric, then we assume the ratio of the two surface gravities can be scaled to unity without loss of generality.} Therefore,
\beq \frac{\d Q}{T} = 2\pi \int T^{M}{}_{\m \n} \ell^\m
\ell^\n (-\bar{\lambda}) d\bar{\lambda} \sqrt{h} d^2 x. \label{PalatiniHeat} \eeq
In the expression above, differently from Eq. (\ref{dQ/T}), we use the affine null vector $\ell^\mu$ with respect to the independent connection and the null geodesic bundle comprising the horizon is now parametrized by $\bar{\lambda}$. On the other hand, as the matter only feels the metric $g_{\mu \nu}$, the relevant volume element is still given by $\sqrt{g}$, reducing to $\sqrt{h}$ on the horizon. In this sense, we can write the relevant horizon volume element as $d \Sigma^\nu  = \ell^\mu \sqrt{h} d^2 x d\bar{\lambda}.$

Along the same parameter $\bar{\lambda}$, we now consider a general variation of the entropy. This has the same form as (\ref{metricdS}), but we express it in a slightly different way,
\beq \delta S = \int \sqrt{h} f \bar{\theta} d^2 x,  \eeq
where
\beq \bar{\theta} = \frac{1}{\sqrt{h}}\frac{d \sqrt{h}}{d \bar{\lambda}}+ \frac{1}{f}\frac{df}{d\lambda} = \frac{d \ln (f \sqrt{h})}{d\bar{\lambda}}. \eeq
This suggests that the new expansion measuring the product of $f$ and the transverse element is the relevant one. Thus, we make the transformation
\beq \sqrt{\bar{h}} = f(\mathcal{R}) \sqrt{h}. \label{transverseconformal} \eeq
so that in terms of this new variable, the entropy is an area entropy
\beq S =\alpha \int \sqrt{\bar{h}} d^2 x. \eeq
Imposing the entropy balance equation and matching order by order in $\bar{\lambda}$ we find the zeroth order equilibrium condition, which can be expressed as
\beq \bar{\theta} = 0  \rightarrow \ell^\mu \bar{\nabla}_\mu (\sqrt{\bar{h}}) =  \ell^\mu \bar{\nabla}_\mu (f' \sqrt{h})  = 0 \label{Palequil}. \eeq
Note that we could have performed the above conformal transformation also in metric $F(R)$ gravity, but in that case the equilibrium condition involves the Levi-Civita connection. On the other hand, the above formula in terms of the independent connection is reminiscent of  the metric compatibility condition (\ref{metriccomp}).

First, note that the vanishing of the expansion $\bar{\theta}$ can be expressed as
\beq \frac{d}{d\bar{\lambda}} \sqrt{det(\bar{g}_{\mu \nu} e^\mu_a e^\nu_b)} = \frac{d}{d\bar{\lambda}} \sqrt{det(\bar{g}_{\mu \nu}) det(e^\mu_a e^\nu_b)} = 0. \eeq
The quantities $e^\mu_a$ are basis vectors in the cross-section of the horizon and the index $a$ runs over the two transverse directions. Since these basis vectors by construction are Lie transported along the horizon, we must have that
\beq \frac{d}{d\bar{\lambda}} \sqrt{-\bar{g}} = 0. \eeq
But since
\beq  \frac{d}{d\bar{\lambda}} \sqrt{-\bar{g}}  = \frac{1}{2} \bar{g}_{\mu \nu} \frac{d}{d\bar{\lambda}} \bar{g}^{\mu \nu} = 0, \eeq
this condition implies that
\beq \frac{d}{d\bar{\lambda}} (\sqrt{-\bar{g}} \bar{g}^{\mu \nu}) = 0. \eeq
Demanding this equation hold for all null vectors $\ell^\mu$, at each point $p$, and using the fact that (\ref{transverseconformal}) implies
\beq \bar{g}_{\mu \nu} = f g_{\mu \nu}, \eeq
we arrive at the metric compatibility equation (\ref{metriccomp}),
\beq \bar{\nabla}_\lambda (f \sqrt{-g} g^{\mu \nu}).   \eeq
Remarkably, the equilibrium condition, which was just an auxiliary condition in the metric formalism, now implies the field equation associated with the variation of the Lagrangian with respect to the independent connection.

Now we go back to the entropy balance law and continue to next order in $\bar{\lambda}$.  We find
\beq \delta S = \alpha \int \sqrt{\bar{h}} \frac{d\bar{\theta}}{d\bar{\lambda}} \bar{\lambda} d^2 x. \eeq
Using the Raychaudhuri equation this can be re-expressed as
\beq \delta S = -\alpha  \int \sqrt{\bar{h}} \mathcal{R}_{\mu \nu} \ell^\mu \ell^\nu \bar{\lambda} d^2 x, \eeq
in terms of the Ricci tensor constructed from $\bar{g}_{\mu \nu}$.  Note the appearance of $\sqrt{h}$ as opposed to the $\sqrt{\bar{h}} = f \sqrt{h}$. Imposing the Clausius relation, and matching both sides, we find
\beq f \mathcal{R}_{\mu \nu} + \Phi g_{\mu \nu} = \left(\frac{2\pi}{\alpha}\right) T^M_{\mu \nu}. \eeq
The $f$ in front of the Ricci tensor has reappeared to account for the mismatch between the effective volume element and the metric volume element felt by matter flux.

In the Palatini formalism, $\bar{\nabla}^\mu T_{\mu \nu} \neq 0$, but the usual metric conservation law $\nabla^\mu T_{\mu \nu} = 0$ holds. Imposing this condition yields
\beq \nabla_\nu \Phi = - f \nabla^\mu \mathcal{R}_{\mu \nu} - \mathcal{R}_{\mu \nu} \nabla^\mu f. \label{conservationlaw}\eeq
The main problem is to calculate the $g$ covariant divergence of the $\bar{g}$ Ricci tensor. We know that $\mathcal{R}_{\mu \nu}$ is related to $R_{\mu \nu}$ via a conformal transformation with conformal factor $f^{1/2}$.  The connection relating the covariant derivatives with respect to the two metrics has the form
\beq \Gamma^{\sigma}{}_{\mu \nu}  = -\delta^{\sigma}_{(\mu} \bar{\nabla}_{\nu)} \ln f' + \frac{1}{2} \bar{g}_{\mu \nu} \bar{g}^{\sigma \delta} \bar{\nabla}_\delta \ln f'. \label{conformalconnection}\eeq
Using the formula (see e.g.~\cite{Koivisto:2005yk})
\beq \nabla^\mu \mathcal{G}_{\mu \nu} = -\nabla^\mu \ln f \mathcal{R}_{\mu \nu} \eeq
and defining $\mathcal{G}_{\m \nu} = \mathcal{R}_{\mu \nu} - \frac{1}{2} g_{\mu \nu} \mathcal{R}$, we ultimately find that
\beq  \nabla_\nu \Phi = -\frac{1}{2} f \bar{\nabla}_\nu \mathcal{R} = -\frac{1}{2} \bar{\nabla}_\nu F, \eeq
so that $\Phi = -1/2 F + const.$ as we expect.  Therefore in the irreversible sector, there is no need for a bulk viscosity term, and the shear viscosity remains the same as in GR and metric $F(R)$ gravity.

\subsubsection{Variation using Levi-Civita connection}
\label{metricvariation}

Now we re-consider the same problem, but working instead with quantities defined with respect to the Levi-Civita connection. Hence we consider the affinely parameterized tangent to be $k^\mu$. The representation of the heat flow and the entropy change is exactly the same as for the metric $F(R)$ case in Section II. As a result, the analysis shows that we have an equation similar to (\ref{testreversible})
\beq  f R_{\m \n}  - \nabla_\m \nabla_\n f + \frac{3}{2f} \nabla_\m f \nabla_\n f + \Phi g_{\mu \nu} = \left(\frac{2\pi}{\alpha}\right) T^{M}{}_{\m \n}, \label{reversiblemet} \eeq
but instead of the fully metric-derived object $f(R)$, now we have $f(\mathcal{R})$. The Ricci tensor that appears explicitly is constructed from the metric and comes from the Raychaudhuri equation in terms of metric compatible variables.

We need to solve for the unknown function $\Phi$. In the metric theory this leads to a contradiction and one had to cancel the kinetic $\nabla_\mu f \nabla^\mu f$ by introducing a bulk viscosity (or equivalently, move it into the irreversible sector), but here we have to consider the presence of two different curvature tensors. Taking a covariant divergence of (\ref{reversiblemet}), we find that
\bea \nabla_\n \Phi =  - (\nabla^\m f) R_{\m \n} - f \nabla^\m R_{\m \n} + \nabla^\m \nabla_\m \nabla_\n f   \nonumber \\
+ \frac{3}{2f^2} \nabla_\m f \nabla^\m f \nabla_\nu f - \frac{3}{2f} \Box f \nabla_\n f \nonumber \\
- \frac{3}{2f} \nabla^\m f \nabla_\m \nabla_\n f.   \eea
Next, we use the Bianchi identity, $\nabla^\m R_{\m \n} = \frac{1}{2} \nabla_\n R$ and a contracted version of the commutator of covariant derivatives
\beq \nabla^\m \nabla_\n V_\m - \nabla_\n \nabla^\m V_\m = R^\tau_\n V_\tau, \label{commutator}\eeq
where $V_\m \equiv \nabla_\m f$ to re-express the second and third terms on the right hand side above. In addition, note that the last term can be re-expressed as
\bea -\frac{3}{2f} \nabla^\m f \nabla_\m \nabla_\n f =  -\frac{3}{4f} \nabla_\n (\nabla^\m f \nabla_\m  f) =\nonumber \\
-\nabla_\n \left(\frac{3}{4f} \nabla^\m f \nabla_\m  f \right) - \frac{3}{4f^2} \nabla_\m f \nabla^\m f \nabla_\nu f. \label{covf}\eea
Combining these results, we obtain
\bea \nabla_\n \Phi = \nabla_\n \left(\Box f-\frac{3}{4f} \nabla_\m f \nabla^\m f \right) - \frac{1}{2} f(\mathcal{R}) \nabla_\n R \nonumber \\
- \left(\frac{3}{4f^2} \nabla_\m f \nabla^\m f + \frac{3}{2 f} \Box f\right) \nabla_\n f. \label{Palconservation}\eea
Here, the key term is the $-1/2 f(\mathcal{R}) \nabla_\n R$. We can introduce the following ansatz
\beq R = \mathcal{R} - Y, \eeq
where $Y$ is some unknown function. Then this term can be manipulated into total derivative terms plus a term multiplying $\nabla_\n f$:
\beq -\frac{1}{2} f(\mathcal{R}) \nabla_\n R = -\frac{1}{2} \nabla_\n f + \frac{1}{2} \nabla_\n (f Y) + \frac{1}{2} Y \nabla_\n f.
\eeq
As total derivatives, the first two pieces contribute to the solution for $\Phi$, which now agrees with the set of terms proportional to $g_{\mu \nu}$ in the single equation of motion (\ref{reducedeom}). The last term above combines with the terms proportional to $\nabla_\n f$ in (\ref{Palconservation}). Demanding that term be zero as a type of consistency or integrability condition implies
\beq Y = \mathcal{R}-R = \frac{3}{2f^2} \nabla_\m f \nabla^\m f + \frac{3}{f} \Box f, \eeq
which is exactly the relationship between the two Ricci scalars in (\ref{RicciSconform}) derived from the conformal transformation.

We have derived (albeit somewhat indirectly) the equations of motion for Palatini $F(\mathcal{R})$ gravity when the connection is eliminated as a independent variable. The fact that equations of motion derived in Sections \ref{Generalvariation} and \ref{metricvariation} are equivalent can be seen \textit{a posteriori} from the conformal relationship between $\bar{g}_{\mu \nu}$ and $g_{\mu \nu}$. This implies that the Killing vector $\chi^\mu$ is geodesic with respect to both the independent and metric connections. Therefore, we have showed that the thermodynamic approach can be extended to encompass the Palatini formalism and Palatini $F(\mathcal{R})$ gravity. In this case no additional bulk viscosity term is needed in the analysis.

\section{Scalar-tensor representations} \label{str}

It is well known that both the metric and Palatini versions of $F(R)$ gravity are equivalent to particular scalar-tensor theories.  First consider the metric $F(R)$ action (\ref{metricaction}). One can treat $f(R) \equiv \frac{\partial F}{\partial R}$ as an auxiliary field $\phi$ and assume $F''(R) \neq 0$ for all $R$. Then one can take the potential $V(\phi)$ as the Legendre transform of $F(R)$ so that $R = V'(\phi)$. Therefore one can rewrite the action in the equivalent form
\beq I_{\omega =0} = \frac{\alpha}{4\pi} \int d^4 x \sqrt{-g} (\phi R + V(\phi) + L_{matt}). \eeq
This is the Jordan frame representation of a Brans-Dicke scalar-tensor theory with the Dicke coupling constant set to $\omega=0$. The corresponding equations of motion are
\bea
R &=& V'(\phi) \\
\phi G_{\mu \nu}&=&\nabla_\mu \nabla_\nu \phi + \left(\frac{2\pi}{\alpha}\right) T^M_{\mu \nu}
-g_{\mu \nu} \Box \phi  \nonumber \\  && - \frac{1}{2} g_{\mu \nu}V(\phi).
\eea
These equations also simply follow from the metric equation of motion (\ref{metricrev}) with the identification $f \equiv \phi$. Hence in the scalar-tensor representation the ``bulk viscosity" term has the form \cite{Chirco:2009dc}
\beq  \delta N_{bulk} =\alpha \int d^2 x d\lambda \lambda  \sqrt{h} \frac{3}{2\phi} k^\mu k^\nu \nabla_\mu \phi \nabla_\nu \phi. \label{scalarflux} \eeq

The procedure is the same for the Palatini action (\ref{Palatiniaction}) and one finds
\beq I_{pal} = \frac{\alpha}{4\pi} \int d^4 x \sqrt{-g} (\phi \mathcal{R} + V(\phi) + L_{matt}).  \eeq
Using the relationship between $\mathcal{R}$ and $R$ found earlier in (\ref{RicciSconform}) we can express this action (up to surface terms) as
\bea I_{\omega = -3/2} = \frac{\alpha}{4\pi} \int d^4 x \sqrt{-g} (\phi R+ \frac{3}{2\phi} \nabla_\mu \phi \nabla^\mu \phi  \nonumber \\
+  V(\phi) + L_{matt}), \eea
which is the Brans-Dicke theory with $\omega = -3/2$.

To apply the thermodynamic formalism, much of the previous analysis can be carried over, just with the identification $\phi = f$. However, it is initially unclear how the equations of motion for the scalar field can emerge out of this analysis. To start, we assume the holographic entropy has the form
\beq S = \alpha \int \sqrt{h} \phi d^2 x. \label{scalarentropy} \eeq
Suppose we follow (\cite{Eling:2006aw}) and cancel out the expansion term by treating it as a part of the irreversible sector. Then we arrive at
\beq  \phi R_{\m \n}  - \nabla_\m \nabla_\n \phi + \Phi g_{\mu \nu} = \left(\frac{2\pi}{\alpha}\right) T^{M}{}_{\m \n}, \label{phireversible} \eeq
for some undetermined function $\Phi$. To determine $\Phi$ we can demand the local conservation of matter-energy as usual. Imagine that we know the action for the matter fields present. It is a functional of  $I_{matt}(g_{\m \n}, \psi)$, where $\psi$ represents some arbitrary matter. Using the diffeomorphism invariance of this action and assuming the matter fields satisfy their equation of motion $\delta I_{matt}/\delta \psi =  0$, one can easily show the following conservation equation holds\footnote{In general, the matter part of the action can also depend on the scalar field: $I_{matt}(g, \psi, \phi)$. Then the matter stress tensor is not conserved: $\nabla^\mu T^{M}{}_{\m \n} = \frac{1}{2} T_{\phi} \nabla_\nu \phi$, where $T_{\phi} = (\sqrt{-g})^{-1} \delta I_{matt}/\delta \phi$.}
\beq \nabla^\mu T^{M}{}_{\m \n} = 0. \eeq

Imposing this equation we find the following equation for $\Phi$,
\beq \nabla_\nu \Phi = - (\nabla^\m \phi) R_{\m \n} - \phi \nabla^\m R_{\m \n} + \nabla^\m \nabla_\m \nabla_\n \phi. \eeq
Using the contracted Bianchi identity and the commutator of covariant derivatives, we are left with
\beq \nabla_\nu \Phi  = \nabla_\n (\Box \phi) - \frac{1}{2} \nabla_\n(\phi R) + \frac{1}{2} R \nabla_\n \phi.  \eeq
Now in order to solve this equation, we must impose the following ``integrability condition" on the last term of the right hand side of the previous equation. Namely, we assume it can be expressed as the derivative of some function, chosen to have the form $\frac{1}{2}~V(\phi)$, i.e. $\frac{1}{2}R\nabla_\n\phi=\frac{1}{2}~\nabla_\n V(\phi)$. Therefore, we have the condition that
\beq \frac{dV}{d\phi} =  R . \label{intcondition}\eeq
Meanwhile, the solution for $\Phi$ is
\beq \Phi= \Box \phi - \frac{1}{2} \phi R + \frac{1}{2} V(\phi) + \Lambda, \eeq
and the reversible equation becomes
\bea \phi R_{\m \n}  - \nabla_\m \nabla_\n \phi - \frac{1}{2} \phi R g_{\m \n} + g_{\m \n} \Box \phi  \nonumber \\
+ \frac{1}{2} g_{\m \n} V(\phi) = \left(\frac{2\pi}{\alpha}\right) T^{M}{}_{\m \n}, \label{metriceom1}\eea
where we have absorbed the cosmological constant $\Lambda$ into the potential $V(\phi)$. This is exactly the set of field equations for the $\omega=0$ theory. The scalar equation of motion is an integrability condition we must impose for consistency with the conservation of local energy-momentum.

Suppose, on the other hand, that we do not introduce a bulk viscosity term. Then equation describing reversible changes is
\beq  \phi R_{\m \n}  - \nabla_\m \nabla_\n \phi + \frac{3}{2\phi} \nabla_\m \phi \nabla_\n \phi + \Phi g_{\mu \nu} = \left(\frac{2\pi}{\alpha}\right) T^{M}{}_{\m \n}. \label{reversible3} \eeq
This has the same form as (\ref{testreversible}), but now that $\phi$ is an independent field, we can repeat the analysis above to solve for the unknown $\Phi$ function and find the scalar equation of motion as an integrability condition.

Taking a covariant divergence of (\ref{reversible3}), we find that
\bea \nabla_\n \Phi =  - (\nabla^\m \phi) R_{\m \n} - \phi \nabla^\m R_{\m \n} \nonumber \\ 
+ \nabla^\m \nabla_\m \nabla_\n \phi + \frac{3}{2\phi^2} \nabla_\m \phi \nabla^\m \phi \nabla_\nu \phi \nonumber \\
- \frac{3}{2\phi} \Box \phi \nabla_\n \phi - \frac{3}{2\phi} \nabla^\m \phi \nabla_\m \nabla_\n \phi.   \eea
As before, we can use the Bianchi identity and the commutator of covariant derivatives (\ref{commutator}) along with the formula (\ref{covf}) in Section (\ref{metricvariation}).
Combining these results, we obtain
\bea \nabla_\n \Phi = \nabla_\n \left(\Box \phi - \frac{1}{2} \phi R -\frac{3}{4\phi} \nabla_\m \phi \nabla^\m \phi \right)  \nonumber \\
+ \left(\frac{1}{2} R - \frac{3}{4\phi^2} \nabla_\m \phi \nabla^\m \phi - \frac{3}{2\phi} \Box \phi  \right) \nabla_\n \phi. \eea
We now impose the integrability condition on the second term as before
\beq \frac{dV}{d\phi} = R+ \frac{3}{2\phi^2} \nabla_\m \phi \nabla^\m \phi - \frac{3}{\phi} \Box \phi, \label{integrability2}\eeq
which allows us to solve for $\Phi$. The resulting metric field equation is
\bea  \phi R_{\m \n}-  \frac{1}{2} \phi R g_{\mu \nu}  - \nabla_\m \nabla_\n \phi + \frac{3}{2\phi} \nabla_\m \phi \nabla_\n \phi \nonumber \\
+ \Box \phi g_{\m \n} - \frac{3}{4\phi} \nabla_\m \phi \nabla^\m \phi g_{\m \n} \nonumber \\
+ \frac{1}{2} V(\phi) g_{\m \n} = \left(\frac{2\pi}{\alpha}\right) T^{M}{}_{\m \n}.
\eea
Therefore we arrive at the equations of motion for Palatini $F(\mathcal{R})$ theory in $\omega = -3/2$ scalar-tensor representation.

\section{General Brans-Dicke theories and ``bulk viscosity" as a heat flux}
\label{GeneralST}

In the above section we showed how the thermodynamic approach can be used to derive the field equations for both metric and Palatini $F(R)$ gravity purely in their scalar-tensor representations. However, the entropy functional (\ref{scalarentropy}) holds also for a general Brans-Dicke theory, which has the action
\bea
I_{gen} = \int \sqrt{-g}\, d^4 x [ \frac{\alpha}{4\pi} (\phi R - \frac{\omega}{\phi} \nabla_\mu \phi \nabla^\mu \phi  \nonumber \\
+ V(\phi) ) +L_{matt} ]. \label{BDaction}
\eea
Previously we were only able to derive the equations of motion for the special cases $\omega=0$ and $\omega= -3/2$, depending on whether ``bulk viscosity" term is needed. In particular, for the $\omega = -3/2$ case equivalent to Palatini, no such term was needed to complete the analysis.

Whether or not we need an additional term appears to be directly related to the existence of an additional propagating scalar degree of freedom in $F(R)$ and scalar-tensor gravity, as was first hypothesized in \cite{Chirco:2009dc}. Since in Palatini $F(\mathcal{R})$  the connection is only an auxiliary field, one would not identify it with any additional propagating degree of freedom. It is possible to more clearly show this distinction between an auxiliary field and a dynamical, propagating one in the scalar-tensor representation. Consider, for example, the $\omega=0$ theory (any general $\omega$ will do). The trace of the metric field equation (\ref{metriceom1}) and the scalar integrability condition (\ref{intcondition}) yield
\beq 3 \Box \phi + 2 V(\phi) -\phi \frac{dV}{d\phi} = \left(\frac{2\pi}{\alpha}\right) T^{M}{}^\mu_\mu,  \eeq
so the propagation of the scalar is determined by the matter sources, as usual. On the other hand, in the special case where $\omega = -3/2$  the same procedure yields
\beq 2 V(\phi) -\phi \frac{dV}{d\phi} = \left(\frac{2\pi}{\alpha}\right) T^{M}{}^\mu_\mu. \eeq
Therefore in this case the scalar field is algebraically related to the matter sources and does not propagate.

We now argue the clear link between the additional flux term associated with the horizon expansion $\theta$ and a propagating scalar degree of freedom indicates that the previous ``bulk viscosity" interpretation in \cite{Eling:2006aw,Chirco:2009dc} was incorrect. First, let us go back and consider the shear squared term. We think of this term as indicating a channel for horizon dissipation, given a gravity theory or equivalently an entropy functional. In GR, this channel is sourced by a flux of gravitational perturbations across the horizon (specifically, a perturbation of the electric part of the Weyl tensor) and gives rise to the Hartle-Hawking tidal heating term \cite{Chandra, Poi:2004, HH, Poi:2005}
\beq \delta N_{shear} = \frac{1}{8\pi G_N} \int \hat{\sigma}_{\mu \nu} \hat{\sigma}^{\mu \nu} dv \sqrt{h} d^2 x. \label{HH}\eeq
The above expression generalizes in metric $F({R})$ in a similar way, acquiring only an overall $f({R})$ factor.

Note that this does not mean gravitational waves are dissipative. While the spacetime dynamics is completely conservative, dissipation only exists for the thermodynamical horizon system.  Furthermore, in the thermodynamical argument, the shear at $p$ only depends on the warping of $B$ and therefore is completely independent of the spacetime geometry.

Gravitational waves have no local stress tensor, and correspondingly the Hartle-Hawking term has a non-local character as the integral over the horizon of an object constructed out of derivatives of the null normal $k^\mu$. This term is consistent with an irreversible flux, which in non-equilibrium thermodynamics \cite{Landau, deGroot} is positive definite and constructed out of derivatives of the state functionals of the system.

It was argued in \cite{Chirco:2009dc} that when the additional scalar degree of freedom is present, there is a new ``gravitational" channel available for dissipating energy (e.g. for relaxing horizon perturbations). However, it is unclear if the related term (\ref{scalarflux}) should be associated to some irreversible branch of the thermodynamic equations.   The problem with this interpretation can be realized by analyzing the form of the ``bulk viscosity" term itself given in (\ref{scalarflux}):
Unlike the shear squared term, it does not depend on derivatives of $k^\mu$ and both the integral over the horizon and the arbitrary $k^\mu$ vectors can be peeled off. Hence, like the other terms with this structure, it has a local interpretation, which is consistent with fact that a scalar field has a local stress tensor. After the $k^\mu$ vectors are peeled off, local terms at $p$ are frame independent. They exist for any observer (accelerated or inertial) in the local patch of spacetime and always end up describing the dynamics of the global spacetime. Indeed, the expansion is no longer an arbitrary quantity defining the local horizon system, but instead fundamentally linked to the derivative of the scalar field on the spacetime. Therefore, if we insist that this term is irreversible, it would imply waves of the scalar field would be dissipative in the spacetime. This is inconsistent with the fact that classical gravitational theories are time reversal invariant.

We now present a new interpretation of this term as a contribution to the heat flux $\delta Q$ of reversible thermodynamics. Let us return to the beginning of the argument and the entropy
\beq S = \int d^4 x \sqrt{h} \phi(x). \eeq
Here we have promoted the entropy density to be an independent field in the spacetime, with dimensions of $[L]^{-2}$. In order to be consistent with the principle of background independence, this field should not be a fixed structure,  it must be varied like other fields. It also must contribute to the total Lagrangian of the theory, i.e.
\beq L_{matt}(g_{\mu \nu}, \psi) + L_{scalar}(g_{\mu \nu}, \phi) \eeq
where $\psi$ represent ``ordinary" matter fields. Upon varying with respect to the metric, we have the usual stress tensor of the various matter fields which do not contribute to the horizon entropy, plus a stress tensor for the scalar field. The components relevant for the heat flux across the horizon are given by contraction with the null vectors
\beq \delta Q \sim k^\mu k^\nu (T^M_{\mu \nu} + T^{\phi}_{\mu \nu}). \eeq
Generally, the Lagrangian for a scalar field consists of possible interaction terms, e.g. a mass squared term, $\phi^4$ term, etc. These can be represented as part of a generic scalar potential $V(\phi)$. However, in the stress tensor, this term's contribution is proportional to $g_{\mu \nu}$, so it does not appear in the heat flux. On the other hand, kinetic terms in the scalar field action must contribute. We assume that the action is constructed out of first derivatives of the scalar field. This eliminates non-minimally coupled k-essence models \cite{ArmendarizPicon:2000dh}, where the scalar fields have non-canonical kinetic terms.

Based on dimensional analysis, the most general contribution of the scalar to the heat flux has the basic form
\beq \delta Q_{scalar} \sim  \frac{\Omega(\phi)}{\phi} k^\mu k^\nu \nabla_\mu \phi \nabla_\nu \phi \eeq
where $\Omega$ is some dimensionless function. Since $\phi$ is dimensionfull, one would have to introduce a new length scale in order to construct a non-trivial $\Omega$. Therefore we take $\Omega$ to be an arbitrary constant. The general form of the heat flux is now
\beq \frac{\delta Q}{T} = - \int d^4 x \sqrt{h} \lambda~ (2\pi T^M_{\mu \nu} k^\mu k^\nu + \left(\frac{\Omega}{\phi}\right) k^\mu k^\nu \nabla_\mu \phi \nabla_\nu \phi). \label{Jordanflux} \eeq
Following the analysis as before, the equation of motion the reversible sector is now (\ref{reversible3}), but with the extra heat flux contribution
\beq  \phi R_{\m \n}  - \nabla_\m \nabla_\n \phi + \left(\frac{3/2-\Omega}{\phi}\right) \nabla_\m \phi \nabla_\n \phi + \Phi g_{\mu \nu} = 2\pi T^{M}{}_{\m \n}, \label{reversiblefinal} \eeq
which captures any Brans-Dicke theory if we set the Dicke constant $\omega= \Omega-3/2$.

In the special case when $\omega=-3/2$, $\Omega=0$ and there is no additional contribution to heat flux from the scalar. This is consistent with the fact that the $\phi$ field is non-propagating (has no kinetic term) in this particular case. In addition, note that when $\omega < -3/2$ the scalar field flux comes in as a ``ghost" with a negative sign and the change in the black hole entropy can longer be positive definite due to the violation of the null energy condition. This is consistent with the study of the classical second law for Brans-Dicke theory done in \cite{Kang:1996rj} and the numerical results of the gravitational collapse of scalar matter pulses \cite{Hwang:2010aj}, which indicated a violation of weak cosmic censorship when $\omega<-3/2$.

It is also interesting to make the transformation to the so-called Einstein frame of the Brans-Dicke theory. One makes a conformal transformation and a redefinition of the scalar field
\bea \tilde{g}_{\mu \nu} &=& \phi g_{\mu \nu} \nonumber \\
d\tilde{\phi} &=& \frac{d\phi}{\phi} \label{Einsteintransform}
\eea
in the action (\ref{BDaction}). The action now has the form
\bea I_{Ein} = \int \sqrt{-\tilde{g}} d^4 x ~ [\tilde{R} - (\omega+3/2) \tilde{\nabla}_\mu \tilde{\phi} \tilde{\nabla}^\mu \tilde{\phi} \nonumber \\
+ \exp(-2 \tilde{\phi}) L_{matt}(\tilde{g})],  \label{Einsteinframe}
\eea
which is just Einstein gravity with the scalar field as a matter field minimally coupled to gravity, but universally coupled to the other matter fields\footnote{Note that our redefinition for the scalar field differs from the standard one used in \cite{valeriobook}. In that case there is an overall factor of $(\omega+3/2)^{-1/2}$ in (\ref{Einsteintransform}), which normalizes the scalar kinetic term to the canonical unity value in the Einstein frame. Hence, in this ansatz one cannot go to the Einstein frame in the $\omega < -3/2$ regime, where the theory is likely to be sick as we discussed earlier. Also in our case one can clearly see that something goes wrong in the same region of parameter space: Eqn.\eqref{Einsteinframe} shows in fact that for $\omega < -3/2$ the kinetic term for the field $\tilde{\phi}$ changes sign leading effectively to a ghost field}.

In the thermodynamic approach, this transformation returns the entropy to just an area in the new conformally related metric. Working in terms of this new metric, we arrive at the Einstein equations
\beq \tilde{G}_{\mu \nu} k^\mu k^\nu = 2\pi T^{total}_{\mu \nu} k^\mu k^\nu, \eeq
where
\beq T^{total}_{\mu \nu} \sim \frac{\delta (I_{scalar}(\tilde{g}, \tilde{\phi})+I_{matt}(\tilde{g},\tilde{\phi},\psi))}{\delta \tilde{g}^{\mu \nu}}. \eeq
In this case there is no need for a bulk viscosity term, but in order to be consistent with the equations of motion we now must explicitly include a scalar flux as a part of the heat flow due to the matter fields. This has the form
\beq \delta Q_{scalar}=  (\omega+3/2) \int d^4 x \sqrt{\tilde{h}} k^\mu k^\nu \tilde{\nabla}_\mu \tilde{\phi} \tilde{\nabla}_\nu \tilde{\phi}. \eeq
Rewriting this term in the Jordan frame using (\ref{Einsteintransform}), we find it is exactly the scalar field flux we argued for in (\ref{Jordanflux}). Therefore the new interpretation of the scalar field contribution as a heat flux ultimately does not depend on the choice of conformal frame. We can work in a frame where the scalar is purely matter, with no contribution to the entropy, or in a frame where it is another gravitational field.

\section{Discussion}

In this paper we have extended the thermodynamics of spacetime formalism to Palatini gravity, where the connection is a priori an independent variable from the metric. We applied this procedure to Palatini $F(\mathcal{R})$ gravity and derived the field equations as a consequence of enforcing an entropy balance law on the local Rindler wedge system. Unlike the metric $F(R)$ case studied previously in \cite{Eling:2006aw}, no ``bulk viscosity" term was required in order to have equations consistent with the conservation of local energy-momentum. Motivated by the fact that both versions of $F(R)$ gravity are equivalent at the classical level to particular Brans-Dicke theories, we considered an entropy density that is a scalar spacetime function $\phi(x)$. This amounts to promoting the inverse of the Newton constant to be an independent scalar field. We showed how the thermodynamic derivation in this case can capture both the field equations of the metric and the scalar field. As a key part of our analysis, we recognized that previous interpretations introducing an irreversible bulk viscosity were incorrect. Instead, we argue that the heat flux $\delta Q$ naturally contains a separate contribution from the scalar field.  This description also is consistent when one works \textit{ab initio} in the Einstein conformal frame of the scalar-tensor theory. In these theories, it seems the bulk viscosity $\xi_B$ should actually be zero.

It is worth noting that if one works a priori the metric $F(R)$ theory, as was done in \cite{Eling:2006aw}, interpreting the extra term needed for consistency with local energy-momentum conservation as a separate heat flux is not as clear. In this representation, there is not a distinct scalar field we need to endow with its own dynamics, only the metric and the general function $f(R)$. Of course, it is well-known that there is an extra dynamical scalar degree of freedom in a theory of gravity which is fourth order in metric derivatives. For example, the trace of the equation of motion (\ref{metricrev}) gives a wave equation relating $\Box f$ to the trace of the matter stress tensor $T^M$; there is no longer just an algebraic link between scalar curvature and $T^M$ as in GR.  Therefore, one can argue that $f$ needs its own dynamics, but it appears there is no way this can be done a priori starting only with an entropy functional $\int \sqrt{h} d^2x  f(R)$.  Instead,  the warning that there is an extra degree of freedom (effectively, an extra flux) is given by the fact that local energy-momentum conservation fails.

In the future, it would be interesting to see if other diffeomorphism invariant theories of gravity admit a thermodynamic interpretation. In particular, what kind of heat fluxes and viscosities appear in different theories? For example, generalized Lovelock gravities are of particular interest and have been studied in a different thermodynamic picture of gravity \cite{Paranjape:2006ca}. Some other interesting examples are the generalized Palatini gravities discussed in \cite{Vitagliano:2010pq} and  the ``metric-affine" theories \cite{Sotiriou:2006qn,Vitagliano:2010sr}, where the matter is now coupled to the independent connection and not just the metric. One can also consider theories with a non-zero torsion, either as a dynamical propagating field \cite{Carroll:1994dq} or algebraically determined by spin, as in the Einstein-Cartan theory \cite{Hehl:1976kj}. How do these types of geometrical structures get mapped into thermodynamics?

Finally, note that while we argued for generalizations of the entanglement entropy density by appealing to less restrictive formulations of the equivalence principle, our choices were always consistent with \cite{Vollick:2007fh,Faraoni:2010yi} Wald's Noether charge entropy formula \cite{Wald:1993nt}. Since the field equations are an assumption in the derivation of the Noether charge entropy, one may worry our approach is just a consistency check. However, a recent work \cite{Brustein:2007jj} has found that the Noether charge entropy applied to static, spherically symmetric spacetimes is equivalent to an area entanglement entropy divided by an effective Newton constant. It would be very interesting to see if this result can be generalized.

\section{Acknowledgements}

We thank V. Faraoni, T. Jacobson, T. Padmanabhan and V. Vitagliano for illuminating discussion and T. Sotiriou for helpful comments on the manuscript.

\end{document}